\documentclass[11pt]{article}
\usepackage{epsfig}
\usepackage{graphicx}
\usepackage{soul}
\usepackage{color}
\usepackage{amsmath}
\usepackage{stackrel}
\begin{document}
\title{Compact Localized States in Engineered Flat-Band $\cal PT$ Metamaterials}
\author{N. Lazarides, G. P. Tsironis \\
        Department of Physics, University of Crete, \\
        P. O. Box 2208, 71003 Heraklion, Greece \\
        National University of Science and Technology ``MISiS'', \\
        Leninsky Prospekt 4, Moscow, 119049, Russia}
\maketitle
\begin{abstract}
The conditions leading to flat (dispersionless) frequency bands in truly 
one-dimensional parity-time ($\cal PT$) symmetric metamaterials comprising 
split-ring resonators (SRRs) arranged in a binary pattern are obtained 
analytically. In this paradigmatic system, in which the SRRs are coupled through 
both electric and magnetic dipole-dipole forces, flat-bands may arise from 
tailoring its natural parameters (such as, e.g., the coupling coefficients 
between SRRs) and not from geometrical effects.
For sets of parameters which values are tailored to flatten the upper band of 
the spectrum, the solution of the corresponding quadratic eigenvalue problem 
reveals the existence of compact, two-site localized eigenmodes. Numerical 
simulations confirm the existence and the dynamic stability of such modes, which 
can be formed through the evolution of single-site initial excitations without 
disorder or nonlinearity.
\end{abstract}

\section*{Introduction}
Considerable research effort has focused the last decades in the development and
investigation of artificial structures such as {\em metamaterials} 
\cite{McPhedran2011,Zheludev2012} and {\em parity-time ($\cal PT-$) symmetric 
materials} \cite{ElGanainy2018}, which exhibit properties not available in 
natural materials. Inspired by Veselago's ideas \cite{Veselago1967}, Pendry and 
his collaborators suggested using split-ring resonator (SRR) arrays 
\cite{Pendry1996} and thin-wire arrays \cite{Pendry1999} to achieve effectively 
negative dielectric permeability and diamagnetic permittivity, respectively, in 
overlapping frequency bands. The combination of these two subsystems into a 
single artificial structure results in a negative refractive index medium, whose 
first realization was made in the turn of the 21st century \cite{Smith2000}. 
The $\cal PT-$symmetric materials originated from the ideas and notions of 
non-Hermitian Quantum Mechanics \cite{Bender2002}, which were later transferred 
to optical lattices \cite{Makris2008} and electronic systems 
\cite{Schindler2012}. The application of these ideas in electronic circuits has 
provided easily accessible experimental configurations as well as a link to the 
electrical circuit picture of SRR-based metamaterials; the latter may acquire 
${\cal PT}$ symmetry which relies on balanced gain and loss 
\cite{Lazarides2013a,Tsironis2014a}. Such ${\cal PT}$ metamaterials (PTMMs) may 
serve as paradigmatic systems that exhibit {\em isolated flat (dispersionless) 
bands}. This is possible because the SRRs in an SRR-based PTMM are coupled 
together both electrically and magnetically through dipole-dipole forces
\cite{Hesmer2007,Sersic2009,Rosanov2011}. That key-property, along with the 
arrangment of the SRRs in a binary pattern, allow for the flattenning of the 
upper band of the two-band frequency spectrum through tailoring the coupling 
coefficients between SRRs. Clearly, this kind of band-flattening {\em is not due 
to geometrical effects}, i.e., the particular lattice structure (one-dimensional 
binary lattice).

Flat energy bands have been observed long ago in the electronic band structure 
of semiconductor heterostructures \cite{White1981} and superconducting cuprates 
\cite{Dessau1993}. Flat-bands were considered in the past as a theoretical 
convenience useful for obtaining exact analytical solutions of ferromagnetism 
(flat-band ferromagnetism) \cite{Tasaki1998}. Recently, the possibility for 
dispersionless (diffraction-free in optics) propagation and robust localization 
in flat-band (FB) systems has initiated intensive research on simple crystal 
structures such as Lieb lattices 
\cite{Bodyfelt2014,Ge2015,Lazarides2017,Slot2017} 
as well as Kagom{\'e} \cite{Kim2017}, merged \cite{Alagappan2016}, cross-stitch 
\cite{Flach2014}, Stub \cite{Real2017}, honeycomb \cite{Wu2007}, rhombic 
\cite{Mukherjee2017}, sawtooth \cite{Zhang2015}, and diamond 
\cite{Leykam2013,Flach2014} lattices that allow for precise FB engineering, even 
in the presence of nonlinearity \cite{Leykam2013} and/or disorder 
\cite{Leykam2013,Nishino2007}. Flat-band engineering methods have been also 
applied in tetragonal lattices beyond the tight-binding picture \cite{Xu2015}. 
Analytical and numerical studies of the spectrum and localization properties of 
Lieb, Kagom{\'e}, and Stub ribbons reveal that $\cal PT$ symmetry, relying on 
gain and loss, does not destroy the flat band in the Lieb ribbon (while it 
destroys the flat-band in the Kagom{\'e} and Stub ribbons) \cite{Molina2015}. 
A detailed account on artificial flat band systems and related experiments is 
given in a recent review article \cite{Leykam2018a}. Furthermore, research on 
numerus diverse systems such as complex networks \cite{Perakis2011}, Weyl 
semimetal superconductors \cite{Lu2015}, organometalic frameworks 
\cite{Liu2013}, twisted bilayer graphene \cite{Cao2018}, graphene grain boundary 
\cite{Dutta2015}, and photonic crystal waveguides \cite{Khayam2009}, has 
also revealed the existence of FBs in their spectrum. Photonic flat bands, which 
have been reviewed in ref. \cite{Leykam2018b}, have been designed for slow light 
propagation \cite{Gersen2005,Li2008}. The existence of one or more FBs in the 
spectrum of a particular system is typically associated with the emergence of 
compact localized eigenmodes. Recently, the first experimental observation of 
diffraction-free propagation of such FB modes has been reported in Lieb photonic 
lattices \cite{Mukherjee2015a,Vicencio2015}, and later in rhombic 
\cite{Mukherjee2017} and Kagom{\'e} \cite{Zong2016} photonic lattices, as well 
as in bipartite optomechanical lattices \cite{Wan2017}.

Here, a truly one-dimensional (1D) PTMM model in which a complete and isolated
FB arises from tailoring its parameters and {\em not from geometrical effects}, 
is presented. A condition for the existence of a FB is obtained analytically, 
which can be satisfied by realistic parameter sets. In the presence of a FB in 
the frequency spectrum, compact, two-site localized eigenmodes are found by 
solving the corresponding quadratic eigenvalue problem (QEP). The existence and 
dynamic stability of such modes is confirmed by numerical simulations.
\begin{figure}[!h] 
\centering
\includegraphics[angle=0, width=0.75 \linewidth]{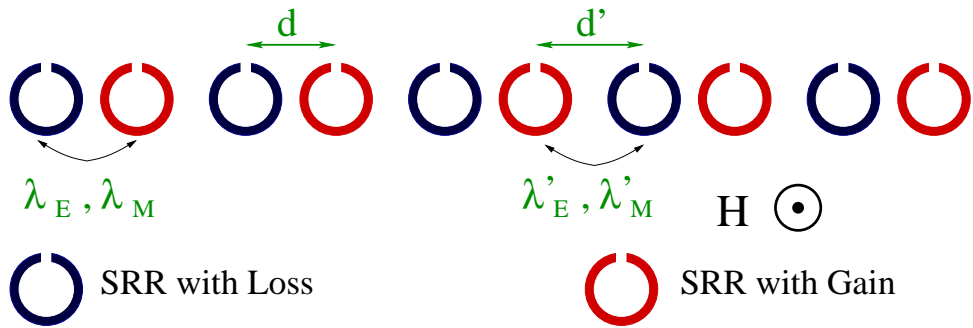}
\caption{Schematic of a $\cal PT$ metamaterial comprising split-ring resonators 
arranged in a one-dimensional binary pattern.
\label{fig1}
}
\end{figure}

\section*{Results}
\subsection*{Modelling, Frequency Dispersion, and Flat-Band Condition}
Consider a 1D array of SRRs arranged in a binary pattern as in Fig. \ref{fig1},
in which the SRRs have alternatingly loss (blue) and gain (red). In a balanced
configuration, which is considered here, the amounts of loss and gain are equal.  
The SRRs shown in Fig. \ref{fig1} (note the mutual orientation of their slits) 
interact both electrically and magnetically through dipole-dipole forces 
\cite{Sersic2009,Rosanov2011}, and can be regarded as $RLC$ circuits, featuring 
a resistance $R$, an inductance $L$, and a capacitance $C$. Gain can be provided 
to an SRR by, e.g., mounting a negative resistance electronic device to its slit
\cite{Lazarides2013a}. Using equivalent circuit models, the normalized equations 
governing the dynamics of the charge $q_n$ stored in the capacitor $C$ of the 
$n$th SRR are obtained as \cite{Lazarides2006,Lazarides2011,Lazarides2013a}
\begin{eqnarray}
\label{1}
   \lambda_M' \ddot{q}_{2n} +\ddot{q}_{2n+1} +\lambda_M \ddot{q}_{2n+2}
   +\gamma \dot{q}_{2n+1} +q_{2n+1} =
   -\{ \lambda_E' q_{2n} +\lambda_E q_{2n+2} \} , \\
\label{2}
   \lambda_M \ddot{q}_{2n-1} +\ddot{q}_{2n} +\lambda_M' \ddot{q}_{2n+1}
   -\gamma \dot{q}_{2n} +q_{2n}  =
   -\{ \lambda_E q_{2n-1} +\lambda_E' q_{2n+1} \} , 
\end{eqnarray}
where $\lambda_E$ and $\lambda_M$ ($\lambda_E'$ and $\lambda_M'$) are 
respectively the electric and magnetic coupling coefficients between SRRs with 
center-to-center distance $d$ ($d'$), $\gamma$ is the gain/loss coefficient 
($\gamma >0$), and the overdots denote derivation with respect to the normalized
temporal variable $\tau =\omega_{LC} t$, with $\omega_{LC}^{-1} =\sqrt{L C}$. 

By substituting the plane wave solution 
\begin{equation}
\label{3}
 q_{2n} = A\, \exp[i( 2 n \kappa -\Omega \tau)], \qquad
 q_{2n+1}=B\, \exp[i( (2 n+1) \kappa -\Omega \tau)]
\end{equation}
into Eqs. (\ref{1}) and (\ref{2}), where $\kappa$ is the normalized wavevector 
and $\Omega$ is the frequency in units of $\omega_{LC}$, and requesting 
nontrivial solutions for the resulting stationary problem, we obtain
\begin{eqnarray}
 \label{5}
  \Omega_\kappa^2 =\frac{1}{2 a} \left(  -b \pm \sqrt{b^2 -4 a c} \right) ,
\end{eqnarray}
where 
\begin{eqnarray}
 a=1 -(\lambda_M^2 +\lambda_M'^{ 2})^2 -2  \lambda_M \lambda_M' \cos(2\kappa),
\nonumber \\
\label{5.2}
 b= 2 \left( \lambda_E \lambda_M +\lambda_E' \lambda_M' \right) 
   +2 \left( \lambda_E \lambda_M' +\lambda_E' \lambda_M \right) \cos(2\kappa)  
   -2+\gamma^2,
\nonumber \\
 c=1 -(\lambda_E^2 +\lambda_E'^{ 2})^2  -2 \lambda_E \lambda_E' \cos(2\kappa).
\end{eqnarray}
In the exact ${\cal PT}$ phase, Eq. (\ref{5}) gives a gapped spectrum with two 
frequency bands separated by a gap. The FB condition is obtained by requesting 
$d\left( \Omega_\kappa^2 \right) / d\kappa =0$ for any $\kappa$ in the first 
Brillouin zone. After tedious calculations, we get
\begin{eqnarray}
\label{7}
 \left[ \frac{1}{\mathcal R'}( 1 -\lambda_M^2 ) +{\mathcal G} +{\mathcal R'} 
        \left( 1 -\lambda_E^2 \right) \right] 
 \left[ \frac{1}{\mathcal R}( 1 -\lambda_M'^2 ) +{\mathcal G'} +{\mathcal R} 
        \left( 1 -\lambda_E'^{ 2} \right) \right] =0 ,
\end{eqnarray}
where 
\begin{equation}
\label{7.2}
 {\mathcal R} =\frac{\lambda_M}{\lambda_E},
\qquad
 {\mathcal R'}=\frac{\lambda_M'}{\lambda_E'},
\qquad
 {\mathcal G} =-2 +\gamma^2 +2 \lambda_E \lambda_M,
\qquad
 {\mathcal G'}=-2 +\gamma^2 +2 \lambda_E' \lambda_M'. 
\end{equation}
\begin{figure}[!t]
\centering
\includegraphics[angle=0, width=0.75 \linewidth]{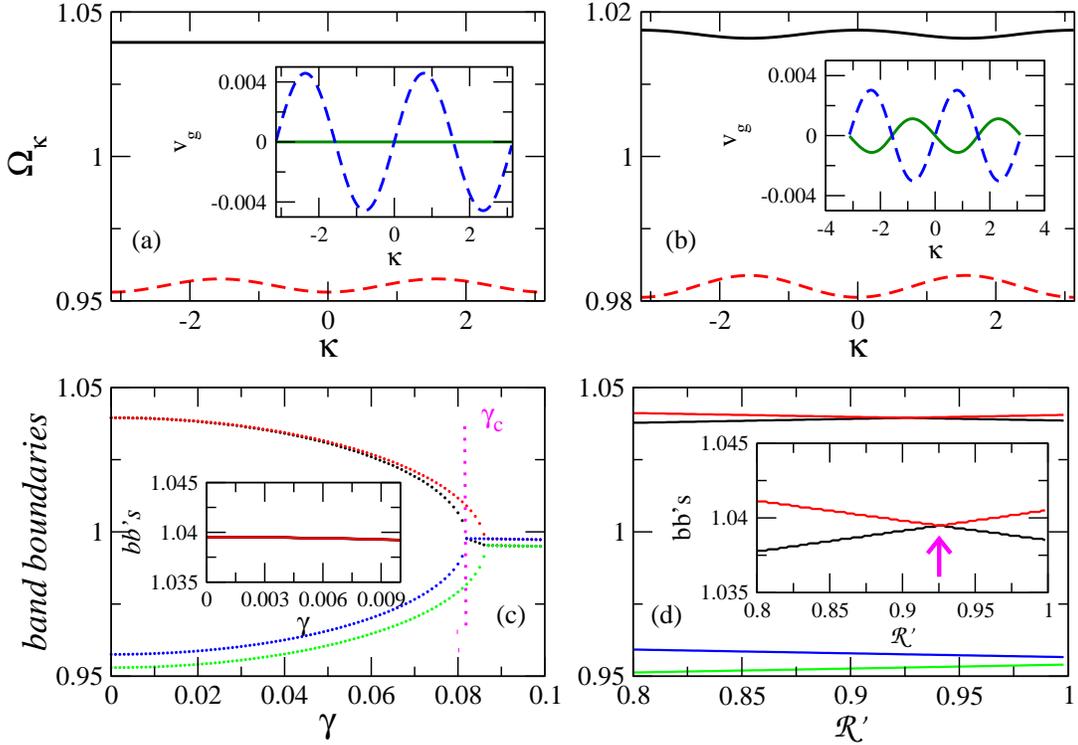}
\caption{
(a) Frequency bands $\Omega_\kappa (\kappa)$ for a ${\cal PT}$ metamaterial with 
    $\gamma=0.003$, $\lambda_E=-0.123952$, $\lambda_M=-0.040128$, 
    $\lambda_E' =-0.027$, and ${\mathcal R'}=0.92547$. 
    The upper band at $\Omega_\kappa \sim 1.03949$ (black-solid curve) is flat. 
    Inset: The corresponding group velocities $v_g$ ($=0$ for the flat band).
(b) Same as in (a) for $\gamma=0.003$, $\lambda_E=-0.055$, $\lambda_M=-0.02$, 
    $\lambda_E' =-0.027$, and ${\mathcal R'}=0.92547$.
(c) The band-boundaries as a function of $\gamma$ for the coupling coefficients 
    in (a). The vertical segment indicates $\gamma =\gamma_c$.
    Inset: Enlargement of the upper band-boundaries for low $\gamma$. 
(d) The band-boundaries as a function of ${\mathcal R'}$, with $\gamma$, 
    $\lambda_E$, $\lambda_E'$, and $\lambda_M$ as in (a). 
    Inset: Enlargement of the boundaries of the upper band around the value of 
    ${\mathcal R'}$ for which the band is flat (shown in (a)).   
\label{fig2}
}
\end{figure}
The values of ${\mathcal R'}$, $\lambda_E$, $\lambda_M$, and $\gamma$, for which 
the expression in the first squared brackets in Eq. (\ref{7}) equals to zero, 
provide a physically acceptable parameter set which flattens the upper band of 
the spectrum over the whole Brillouin zone. For such a parameter set, the 
calculated frequency spectrum (from Eq. (\ref{5})) contains a {\em completely 
flat, isolated upper band} (Fig. \ref{fig2}(a)) with zero group velocity $v_g$ 
for any $\kappa$ in the first Brillouin zone (inset).

The electric and magnetic coupling coefficients between single SRRs depend 
crucially on their mutual position. In the model adopted here, the SRRs assume 
planar geometry; even in this case, however, the coupling coefficients depend 
strongly on the mutual orientation of the gaps. For the configuration of Fig. 
\ref{fig1}, these coefficients have been calculated accurately and plotted in 
figure 1(a) of ref. \cite{Rosanov2011} using the approach presented in ref. 
\cite{Powell2010}, where the effect of retardation has been also taken into 
account. In these works, single SRRs having typical dimensions have been 
considered, while the authors have verified that the resonance frequencies of 
pairs of SRRs coupled with the calculated coefficients match the resonances 
found by direct numerical simulations using commercial software (CST Microwave 
Studio). This indicates that the calculated coupling coefficients can 
quantitatively describe the near-field interaction between SRRs.
According to ref. \cite{Rosanov2011}, the coupling coefficients used in Fig. 
\ref{fig2}(a) correspond to distances $d \sim 2$ and $d' \sim 3.7$ between 
neighboring SRRs in units of their radii, that seems in principle feasible. 
On the contrary, for parameter 
sets not satisfying Eq. (\ref{7}), non-flat bands with $\kappa-$dependent group 
velocities $v_g$ such as those shown in Fig. \ref{fig2}(b) are typically 
obtained. The band-boundaries (BBs), i.e., the extremal frequencies in each 
band, are shown as a function of the gain/loss coefficient $\gamma$ in Fig. 
\ref{fig2}(c). A critical value of that coefficint, $\gamma =\gamma_c$, 
separates the exact or unbroken from the broken $\cal PT$ phase. For 
$\gamma < \gamma_c$ (exact or unbroken $\cal PT$ phase), the BBs of the upper 
band (red and black dotted curves) practically coincide for a substantial 
interval of $\gamma$ indicating zero bandwidth (i.e., a FB), while the width of 
the lower band (limited by the blue and green dotted curves) remains almost 
constant up to $\gamma =\gamma_c$. The flatness of the upper band for low 
$\gamma$ can be seen more clearly in the inset. In Fig. \ref{fig2}(d), the BBs 
are plotted as a function of ${\mathcal R'}$; the width of the upper band 
(limited by the red and black solid curves) consecutively decreases, goes 
through zero at a critical value of ${\mathcal R'}$ indicated by the arrow, and 
then increases with increasing ${\mathcal R'}$ (see also the inset). At that 
critical value of ${\mathcal R'}$ the two bands coincide with those shown in 
Fig. \ref{fig2}(a). 

\begin{figure}[!t]
\includegraphics[angle=0, width=0.75 \linewidth]{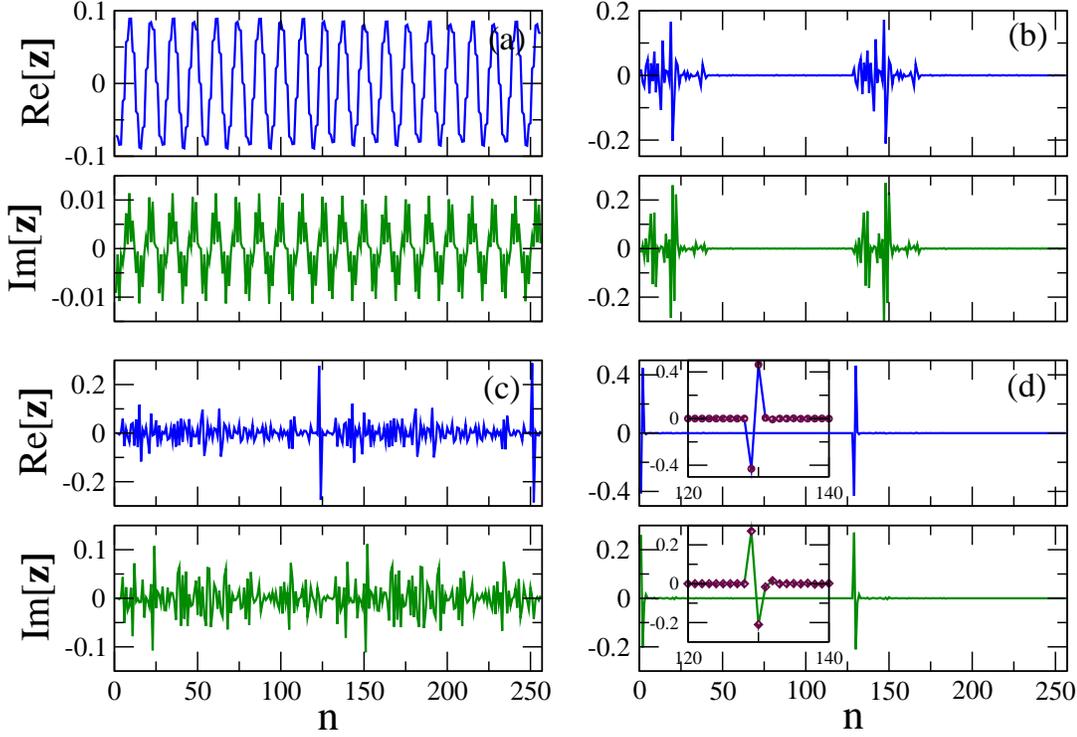} 
\caption{
  Real (blue) and imaginary (green) parts of four eigenmodes obtained by solving
  the quadratic eigenvalue problem Eq. (\ref{10}) for $N=128$, $\gamma=0.01$, 
  $\lambda_E =-0.1200046$, $\lambda_M=-0.0400493$, $\lambda_E' =-0.027$, and 
  ${\mathcal R'}=0.9291948$. 
  The eigenfrequency of the flat band is $\Omega_{FB} \simeq 1.03740$.
(a) An extended eigenmode from the lower (non-flat) band.
(b) A partially localized flat-band eigenmode.
(c) A highly localized flat-band eigenmode.
(d) The compact, two-site localized flat-band eigenmode.
  Insets: Enlargements around the localization region.
\label{fig3}
}
\end{figure}

\subsection*{The Quadratic Eigenvalue Problem}
Eqs. (\ref{1}) and (\ref{2}) can be written in matrix form as 
\begin{eqnarray}
\label{8}
 \hat{\bf{M}} \ddot{\bf Q} =\hat{\bf{C}}_1 \dot{\bf {Q}} +\hat{\bf{K}} {\bf Q},
\end{eqnarray}
where ${\bf Q} =[q_1 ~q_2  ~...  ~q_N]^T$ is an $N-$dimensional vector, with $N$
being the total number of SRRs, and the $N\times N$ matrices $\hat{\bf{M}}$ and 
$\hat{\bf{K}}$ are given by
\begin{eqnarray} 
\label{s04}
\hat{\bf M}=
  \begin{bmatrix}
    1         & \lambda_M & 0          & 0           & 0          & 0          & 0          & \hdots \\
    \lambda_M & 1         & \lambda_M' & 0           & 0          & 0          & 0          & \hdots \\
    0         & \lambda_M'& 1          & \lambda_M   & 0          & 0          & 0          & \hdots \\
    0         & 0         & \lambda_M  & 1           & \lambda_M' & 0          & 0          & \hdots \\
    0         & 0         & 0          & \lambda_M'  & 1          & \lambda_M  & 0          & \hdots \\
    0         & 0         & 0          & 0           & \lambda_M  & 1          & \lambda_M' & \hdots \\
    0         & 0         & 0          & 0           & 0          & \lambda_M' & 1          & \hdots \\
    \vdots    & \vdots    & \vdots     & \vdots      & \vdots     & \vdots     & \vdots     & \ddots 
  \end{bmatrix} , ~~~
\hat{\bf K}=(-1)
  \begin{bmatrix}
    1         & \lambda_E & 0          & 0           & 0          & 0          & 0          & \hdots \\
    \lambda_E & 1         & \lambda_E' & 0           & 0          & 0          & 0          & \hdots \\
    0         & \lambda_E'& 1          & \lambda_E   & 0          & 0          & 0          & \hdots \\
    0         & 0         & \lambda_E  & 1           & \lambda_E' & 0          & 0          & \hdots \\
    0         & 0         & 0          & \lambda_E'  & 1          & \lambda_E  & 0          & \hdots \\
    0         & 0         & 0          & 0           & \lambda_E  & 1          & \lambda_E' & \hdots \\
    0         & 0         & 0          & 0           & 0          & \lambda_E' & 1          & \hdots \\
    \vdots    & \vdots    & \vdots     & \vdots      & \vdots     & \vdots     & \vdots     & \ddots 
  \end{bmatrix}. 
\end{eqnarray} 
The $N\times N$ matrix $\hat{\bf{C}}_1$ is diagonal, with 
$(\hat{\bf{C}}_1)_{n,n} =\gamma (-1)^{n}$ ($n=1,...,N$). By substituting
${\bf Q} = {\bf Q}^{e} \, e^{i \Omega \tau}$ into Eqs. (\ref{8}) we get 
\begin{equation}
\label{9}
  \left\{ \Omega^2 \hat{\bf M} +\Omega \hat{\bf C} +\hat{\bf K} \right\} 
        {\bf Q}^{e} =\hat{\bf 0} , 
\end{equation}
where $\hat{\bf C} \equiv i\hat{\bf C}_1$. Eq. (\ref{9}) is a QEP 
\cite{Tisseur2001} with $\Omega$ being the eigenvalue and ${\bf Q}^{e}$ the 
corresponding $N-$dimensional eigenvector. It can be solved by standard 
eigenproblem solvers after its {\em linearization by the classical augmentation 
procedure} \cite{Duncan1935,Afolabi1987}, which transforms square matrices of 
order $N$ to $2 N$. Then, Eq. (\ref{9}) is reduced to a standard eigenvalue 
problem (SEP) 
\begin{equation}
\label{10}
   \hat{\bf D} {\bf z} =\Omega {\bf z}, ~~
   {\bf z} = \begin{bmatrix}
      {\bf Q}^{e}  \\
      \Omega {\bf Q}^{e}         
   \end{bmatrix}, ~~
   \hat{\bf D} = \begin{bmatrix}
      \hat{\bf 0}                  &  \hat{\bf I}   \\
     -\hat{\bf M}^{-1} \hat{\bf K} & -\hat{\bf M}^{-1} \hat{\bf C} 
   \end{bmatrix},
\end{equation}
where $\hat{\bf I}$ is the $N\times N$ identity matrix and $\hat{\bf M}^{-1}$ 
the inverse of $\hat{\bf M}$. In what follows, $\gamma <\gamma_c$ so that the 
system is in the exact (unbroken) $\cal PT$ phase, and therefore all its 
eigenvalues are real.
A few selected eigenmodes ${\bf Q}^{e}$ are shown in Fig. \ref{fig3}. The 
eigenmodes from the non-flat (lower) band are all extended, as that shown in 
Fig. \ref{fig3}(a). The other three eigenmodes in Fig. \ref{fig3} belong to the 
flat (upper) band, and therefore they correspond to the same frequency 
eigenvalue $\Omega_\kappa =\Omega_{FB}$. Figs. \ref{fig3}(b), (c), and (d) show 
a partially localized, a highly localized, and a compact two-site localized 
eigenmode, respectively. The insets in Fig. \ref{fig3}(d) enlarge the 
localization region. The existence of compact localized eigenvectors in a class 
of FB models has been also demonstrated in ref. \cite{Flach2014}. 
In that work, 
a model with at least one FB was considered, whose eigenvectors in the Bloch 
representation may be mixed to obtain highly localized FB eigenvectors, due to 
macroscopic degeneracy. While there is not any general theorem which states that 
among all these combinations there will be compact localized eigenvectors, such 
eigenvectors do exist in several FB models. Compact localized states were 
constructed, which are actually exact FB localized eigenstates, and then 
classified according to the number of unit cells they occupy. Such compact 
localized states occupying one unit cell form a complete and orthogonal basis 
that makes possible to detangle them from the rest of the lattice. The inversion 
of the detangling procedure provides the most general FB generator having 
localized eigenstates that occupy one unit cell. Such models include cross-stich
and diamond chains, both quasi-one dimensional at the single-site level.
Here, a purely one-dimensional system is considered in the form of a $\cal PT$
symmetric SRR chain arranged in a binary pattern, whose geometry does not in 
general provides a FB. However, that system does possess a FB when the 
coupling coefficients and the gain/loss factor satisfy the condition Eq. (\ref{7}).
In that case, the resulting QEP is directly solved after being transformed into
a SEP, demonstrating the existence of a compact localized FB eigenstate as that 
shown in Fig. \ref{fig3}(d).   
In the light 
of these findings, it would be expected that the PTMM in the exact $\cal PT$ 
phase supports compact localized excitations which in general could be expressed 
as linear superpositions of a small number of eigenmodes.

The existence of localized states or modes in discrete lattices has been
investigated intensively in the past. It has now been established that
it may be due to quenched disorder (random lattices), a phenomenon based on
wave interference which is known as Anderson localization \cite{Anderson1958}.
In the presence of randomly distributed impurities in a metal, for example,
different paths taken by an electron can interfere destructively, leading to
localization of its wavefunction. The concept of Anderson localization is
applicable to, and has actually been observed experimentally in a variety of
physical systems. Localization may also appear in nonlinear lattices which are
perfectly periodic; this effect is known as intrinsic localization, which leads
to localized states of the discrete breather type \cite{Flach2008}.
Discrete breather excitations can be created in nonlinear lattices from an
initially extended state through the standard modulational instability
mechanism. Such localized states have been also experimentally observed in a
variety of physical systems.
Apart from the case of quenched disorder and nonlinearity, localized states
states can also arise in tight-binding (nearest-neighbor) models from particular
lattice geometries in which destructive interference leads to the emergence of
FBs. FB models typically possess localized eigenstates, although
there is no general theorem to guarantee their existence. Since. in the model
considered here, neither disorder nor nonlinearity is present, the apearance 
of compact localized eigenstates for the PTMM is only due to the existence of
the FB in its frequency spectrum.

\begin{figure}[!t]
\includegraphics[angle=0, width=0.75 \linewidth]{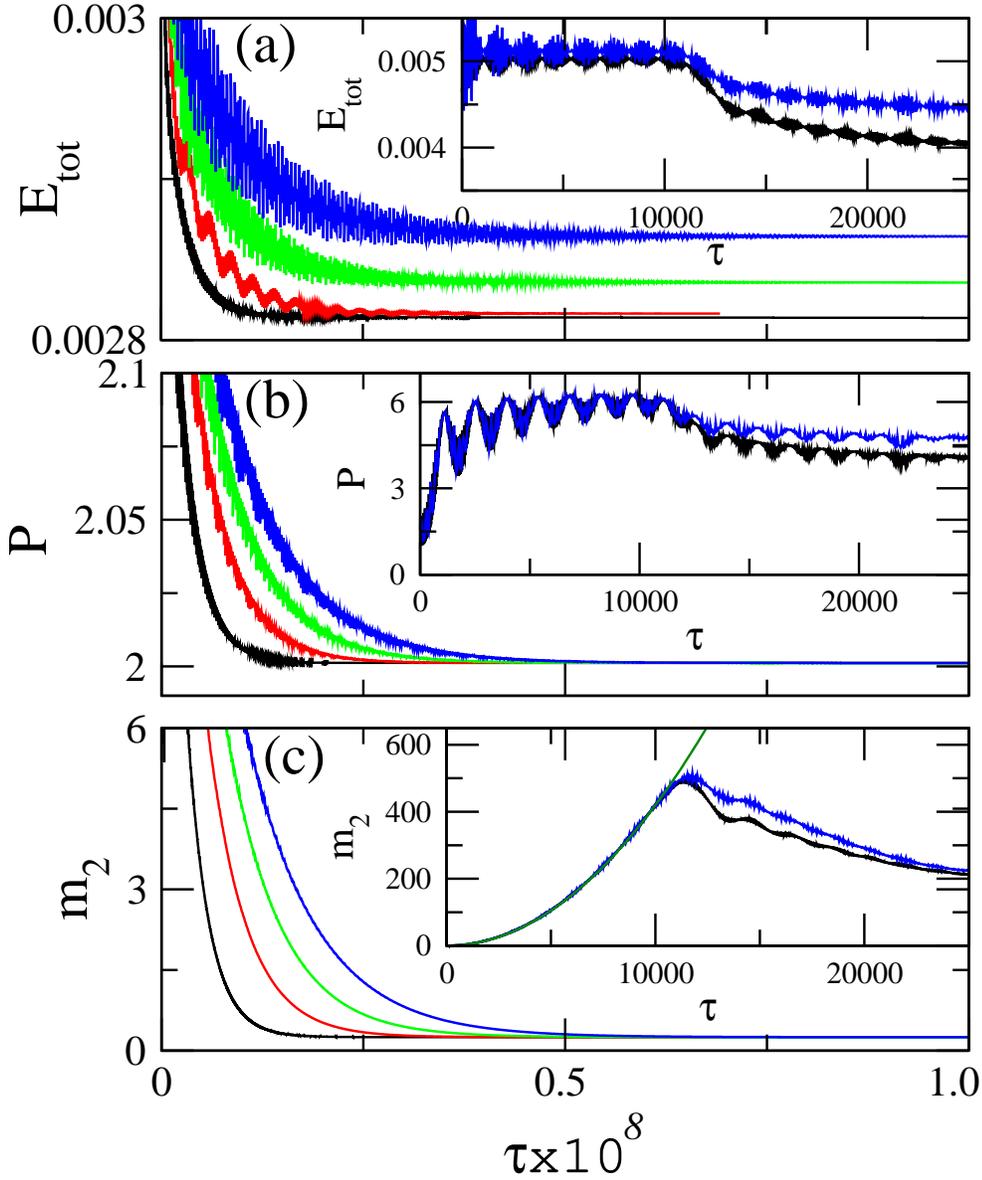}
\caption{
 Total energy $E_{tot}$ (a), energetic participation number $P$ (b), and second 
 moment $m_2$ (c), as functions of $\tau$ for $N=128$, $N_e =16$, and
     $\gamma=0.003$, $\lambda_E=-0.123952$,  $\lambda_M=-0.040128$,  
     ${\mathcal R'}=0.92547$ (black); 
     $\gamma=0.006$, $\lambda_E=-0.120021$,  $\lambda_M=-0.0400518$, 
     ${\mathcal R'}=0.9288096$ (red);
     $\gamma=0.009$, $\lambda_E=-0.1200284$, $\lambda_M=-0.0399891$, 
     ${\mathcal R'}=0.9290096$ (green);
     $\gamma=0.012$, $\lambda_E=-0.120122$,  $\lambda_M=-0.0400118$, 
     ${\mathcal R'}=0.9293204$ (blue).
 Insets: The curves for $\gamma=0.003$ (black) and $0.012$ (blue) in a short 
 time-scale. 
 In the inset in (c), the curve $4.2\times 10^{-6} \tau^2$ (green) is also 
 plotted.
\label{fig4}
}
\end{figure}

\subsection*{Numerical Simulations}
Eqs. (\ref{1}) and (\ref{2}), implemented with free-end boundary conditions 
$q_0 (\tau) =q_{N+1}(\tau) =0$, are integrated in time with a standard $4$th 
order Runge-Kutta algorithm. The initial conditions are single-site excitations 
of the form 
\begin{eqnarray}
\label{10.2}
  q_{n=N/2+1} (\tau=0) =A ~~{\rm for}~~n =N/2+1, 
\qquad
  q_{n=N/2+1} (\tau=0) =0 ~~{\rm for}~~n \neq N/2+1, \\ 
\label{10.3}
  \dot{q}_n (\tau=0) =0 ~~{\rm for}~~n=1,...,N, 
\end{eqnarray}
where typically $A=0.1$. For $N_e$ SRRs at each end of the PTMM, gain has been 
replaced by equal amount of loss, as if the $\cal PT$ metamaterial were embedded 
into a lossy metamaterial. It is found empirically that this is the most 
effective way to stabilize localized modes in PTMMs, even in the presence of 
nonlinearity \cite{Lazarides2013a} (in which case the localized modes are of the 
discrete breather type). The need for introducing these lossy parts at each end 
of the PTMM comes from the fact that the initial condition Eq. (\ref{10.2}) is 
clearly not an exact eigenmode of Eqs. (\ref{1}) and (\ref{2}). However, after 
long time-integration, that type of initial condition may relax to an exact 
eigenmode with a lower energy than the initial one. During this process, some of 
the initial energy spreads towards the ends of the PTMM, where it is dissipated.
Thus, these lossy parts help the excess energy to go smoothly away during the 
transient phase of integration, and allow the formation of a stable compact 
localized FB state with constant energy. This gradual energy removal is clearly 
observed in Fig. \ref{fig4}(a) discussed below.
\begin{figure}[!t]
\includegraphics[angle=-90, width=0.75 \linewidth]{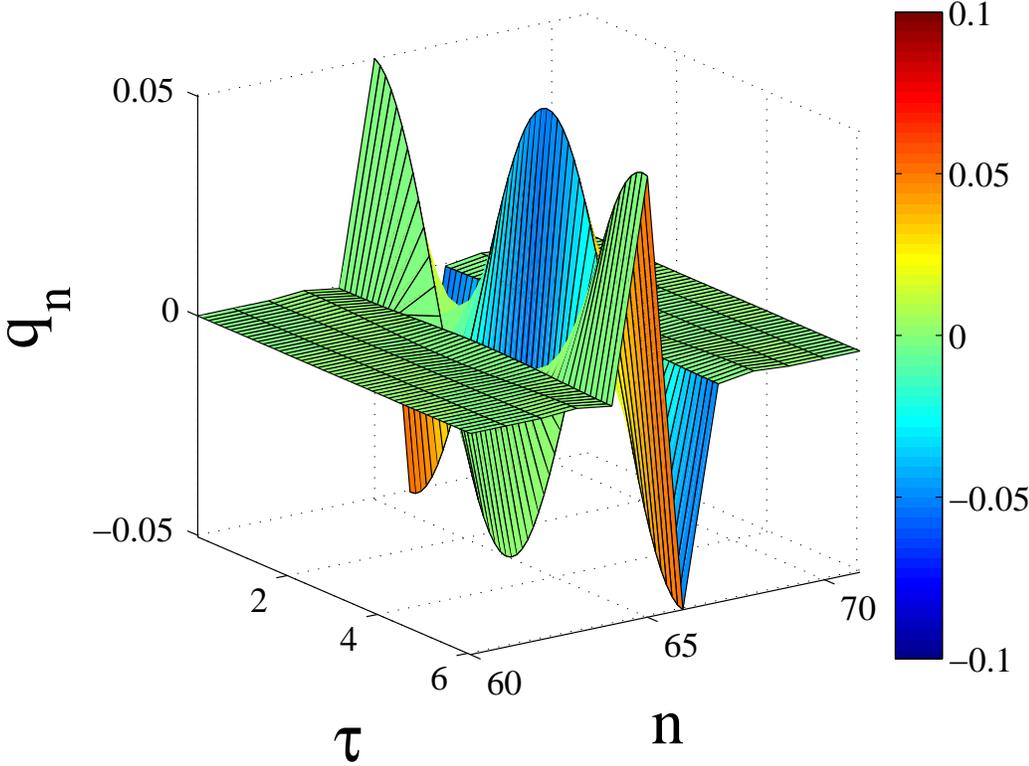} 
\caption{
Spatiotemporal diagram of the charges $q_n$ after $\sim 10^8$ time units of 
integration for $\gamma=0.01$, $\lambda_E=-0.1200046$, $\lambda_M=-0.0400493$, 
${\mathcal R'}=0.9291948$, $N=128$, $N_e =16$ ($\Omega_{FB} \simeq 1.03740$).
Only part of the $\cal PT$ metamaterial is shown for clarity.
\label{fig5}
}
\end{figure}

The total energy, the (energetic) participation number, and the second moment 
\begin{eqnarray}
\label{11}
  E_{tot}= \frac{1}{2} \left\{ 
           \dot{\bf Q}^T \dot{\bf Q} +{\bf Q}^T {\bf Q}
+{\bf Q}^T \hat{\bf{K}} {\bf Q} +\dot{\bf Q}^T \hat{\bf{M}} \dot{\bf Q} \right\},
\qquad
  P=\frac{1}{\sum_n \epsilon_n^2},
\qquad
  m_2 =\sum_n (n -\bar{n})^2 \, \epsilon_n,
\end{eqnarray}
respectively, where $\bar{n} =\sum_n n \, \epsilon_n$ is the center of energy 
with $\epsilon_n =E_n / E_{tot}$ being the energy density, are shown in Fig. 
\ref{fig4} as functions of $\tau$. The four curves are obtained for parameter 
sets which satisfy the FB condition; the coupling coefficients have very similar 
values, while $\gamma$ increases from $0.03$ to $0.12$ in steps of $0.03$. 
In all cases, a steady FB state is reached at the end of the integration time; 
however, the transient period is longer and $E_{tot}$ is higher for 
higher $\gamma$. In the insets of Figs. \ref{fig4}(a), \ref{fig4}(b), and 
\ref{fig4}(c), the quantities $E_{tot}$, $P$, and $m_2$, respectively, are 
plotted for the first $25,000$ time units (t.u.). The energy $E_{tot}$ strongly 
fluctuates, but it remains on average constant until $\tau \sim 12,000$ t.u. 
At the same time, $P$ increases while fluctuating strongly and reaches $P=6$, 
while $m_2$ increases $\propto \tau^2$ indicating ballistic spreading of the 
initial state. Around $\tau \sim 12,000$ t.u., the energy which spreads out from 
the initial state has reached the lossy ends of the PTMM where it is dissipated. 
After that time, $E_{tot}$, $P$, and $m_2$ decrease gradually until they 
saturate to constant values with vanishingly small fluctuations, indicating that 
a steady state has been reached. The constancy of $m_2$, in particular, 
indicates that energy spreading has been stopped. Part of the initial excitation 
remains localized on two SRRs, as indicated by the value of $P =2$ which 
measures the number of the energetically strongest excited sites. 

\begin{figure}[!t]
\includegraphics[angle=0, width=0.75 \linewidth]{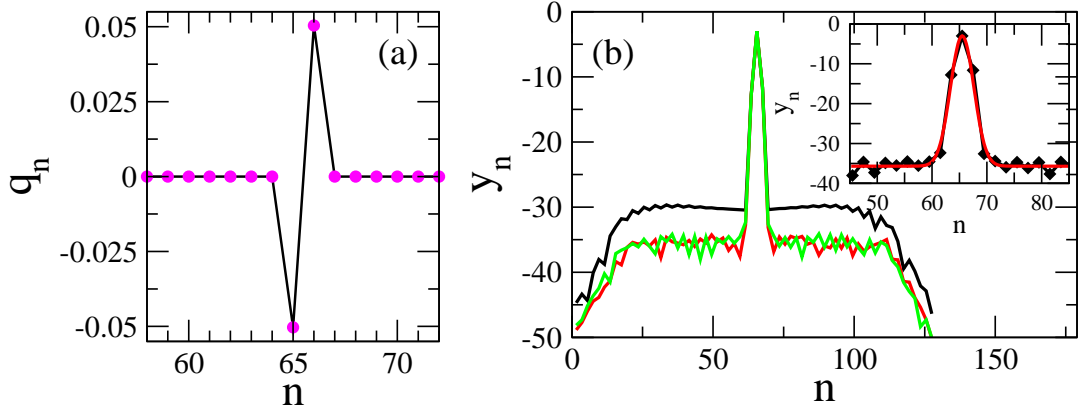}
\caption{
(a) Charge profile $q_n$ at maximum oscillation amplitude. Parameters as in 
    Fig. \ref{fig5}.
(b) The function $y_n$ after integration for $\tau_0 =1.5\times 10^6$ (black), 
    $2\times 10^6$ (red), $3\times 10^6$ (green), time units. 
    Inset: Fitting of $y_n$ with an appropriate function (see text)
\label{fig6}
}
\end{figure}

Inspection of the $q_n (\tau)$ in Fig. \ref{fig5} reveals that the excitation is 
localized to only two, neighboring SRRs. The charges stored in the capacitors of 
these SRRs oscillate in anti-phase with frequency $\Omega_{FB}$; charge 
oscillations in the rest of the SRRs seems to have vanishingly small amplitude 
(see also Fig. \ref{fig6}(a)). Note that if the FB condition Eq. (\ref{7}) is 
not satisfied, the initial excitations disperse rapidly in the lattice without 
leaving behind any trace of localization. Compact localized 
excitations, which tails decay as a stretched exponential or superexponential, 
often appear in discrete systems with nonlinear dispersion 
\cite{Eleftheriou2000}. Here, the compact localized states appear 
solely due to the FB in the absence of nonlinearity or disorder, and hence there 
are significant differences. In Fig. \ref{fig6}(b), the quantity 
$y_n =\ln\left[ \frac{1}{2} \left( |q_{2n-1}| +|q_{2n}| \right) \right]$ is 
plotted as a function of $n$; the three curves correspond to different
integration times, $\tau_0$. In obtaining the results in Fig. \ref{fig6}(b),
convergence to a steady localized FB state was accelerated by continually
eliminating the energy spreading away from the localization region during half
of the integration time, i.e., for $\tau_0 /2$ time units. During this time,
we set $q_n =\dot{q}_n =0$ for $n=1,..., N/2-3$ and $n=N/2+6,...,N$, every $10$ 
periods $T_{FB} =2\pi/\Omega_{FB}$ of integration. This does not affect the 
region in which we expect the compact localized FB state to be formed, i.e., 
the eight sites for 
$n=\frac{N}{2}-2, \frac{N}{2}-1, ... , \frac{N}{2}+4, \frac{N}{2}+5$ 
(localization region). Note that for the chosen initial condition, the compact 
two-site FB localized state is generated at the sites $n=N/2+1$ and $n=N/2+2$, 
i.e., in the middle of the localization region. For $\tau > \tau_0 /2$ the 
integration proceeds without elimination of the energy for $\tau_0 /2$ more time 
units. 
The almost horizontal segments between $16 < n < 63$ and 
$66 < n < 112$ (i.e., those parts of the lattice which do not 
belong either to the lossy ends nor to the localization region) correpond to 
the tails of the localized FB states. With increasing $\tau_0$, the tails 
become more and more negative until they saturate at $y_n \sim -35$ (the limit 
of double precision arithmetics) at $\tau_0 \sim 2\times 10^6$. In the inset, 
the $y_n$ profile at $\tau_0 =3\times 10^6$ t.u. (black curve) is fitted by 
$y_n =b +(\alpha_0 -b) \exp\left[ c (n-x_0)^2 \right]$ (red curve), where 
$x_0=65.5$ and $\alpha_0 =-2.9887$ are taken from the numerical data. The 
fitting parameters are $b=-35.684$ and $c=-0.109955$. 
Using the procedure of energy elimination, the relaxation time towards the
formation of a compact localized FB state is reduced considerably. For example,
after $3\times 10^6$ time units of time-integration, the quantities $E_{tot}$,
$P$, and $m_2$ take respectively the values $2.855\times 10^{-3}$, $2.005$, and
$2.826\times 10^{-1}$. In order to reach these values without energy elimination,
the integration time needed is respectively $\sim 9.13 \times 10^6$, 
$\sim 2.82\times 10^7$, and $\sim 4.62 \times 10^7$ time units. Note that 
without energy elimination there are stil significant fluctuations around the 
above values for $E_{tot}$, $P$, and $m_2$. Thus, the quantities $E_{tot}$, $P$, 
and $m_2$ do not seem to relax at the same rate when they are calculated with 
and without elimination of energy. However, as far as the localization region is 
concerned, a comparison of the relaxation times for $P \simeq 2.005$, i.e., 
$3\times 10^6$ and $\sim 2.82\times 10^7$, indicates a difference of an order of 
magnitude. Note that $P \simeq 2.005$ indicates that there are only two strongly 
excited sites in the PTMM. 

\section*{Conclusion}
By tailoring the model parameters of a 1D PTMM comprising SRRs arranged in a 
binary pattern, an isolated and completely FB may appear in its frequency 
spectrum. The analytical condition, which those parameters have to satisfy in 
order to flatten the upper band, is obtained. The solution of the QEP reveals 
the existence of compact, localized FB eigenmodes, in which most of the energy 
is concentrated in two neighboring SRRs which are separated by distancce $d'$. 
This is consistent with earlier results on nonlinear binary PTMMs, for which the 
fundamental discrete breathers are two-site ones (and not single-site ones) 
\cite{Lazarides2013a}. The formation of compact, two-site localized FB states 
from single-site initial excitations is numerically confirmed. Since the FB is 
isolated, these FB states could be continued into compact breather-like 
(nonlinear) excitations. Note that the possibility of flattening one of the 
bands of the spectrum is solely due to model parameter engineering, and not to 
any geometrical effects. In this aspect, the existence of two types of coupling 
between SRRs, i.e., electric and magnetic coupling, is crucial. Note also that 
similar results would have been obtained for $\gamma =0$, in which case only 
three parameters, i.e., $\lambda_E$, $\lambda_M$, and ${\mathcal R'}$, should be 
matched to satisfy the corresponding FB condition. However, as it is 
demonstrated here, the band-flattening capability is not harmed by the $\cal PT$ 
symmetry, as long as the PTMM is in the exact $\cal PT$ phase.



\section*{Acknowledgements}
 This work was partially supported by the Ministry of Science and Higher 
 Education of the Russian Federation in the framework of Increase 
 Competitiveness Program of  NUST  «MISiS» (No. K2-2017-006), implemented by a 
 governmental decree dated 16th of March 2013, N 211. 
 This research has been financially supported by General Secretariat for 
 Research and Technology (GSRT) and the Hellenic Foundation for Research and 
 Innovation (HFRI) (Grant No: 203). \\
 NL gratefully acknowledges the Laboratory for Superconducting Metamaterials, 
 National University of Research and Technology ``MISiS'' for its warm 
 hospitality during visits.
\section*{Author contributions statement}
 N.L. performed the numerical simulations.
 N.L. and G.P.T. developed the theory, analyzed the results and wrote the paper.
\section*{Additional information}
\noindent {\bf Competing interests:} 
 The authors declare no competing interests.
\end{document}